\documentclass[final]{ias2}

\usepackage{amsmath} 
\usepackage{graphicx} 
\usepackage{multirow}
\usepackage{array} 
\usepackage{textcomp}

\usepackage{hyperref} 

\begin{document}

\markboth{Low Energy Neutrino Measurements}{D. D'Angelo}

\title{Low Energy Neutrino Measurements}

\author[mi]{Davide D'Angelo} 
\email{davide.dangelo@mi.infn.it}
\address[mi]{Universit\`a degli Studi di Milano e I.N.F.N. sez. di Milano - via Celoria, 16 - 20133 Milano (Italy)}

\begin{abstract}
Low Energy solar neutrino detection plays a fundamental role in understanding both solar astrophysics and particle physics.
After introducing the open questions on both fields, we review here the major results of the last two years and expectations for the near future from Borexino, Super-Kamiokande, SNO and KamLAND experiments as well as from upcoming (SNO+) and planned (LENA) experiments.
Scintillator neutrino detectors are also powerful antineutrino detectors such as those emitted by the Earth crust and mantle. 
First measurements of geo-neutrinos have occurred and can bring fundamental contribution in understanding the geophysics of the planet.
\end{abstract}

\keywords{Solar Neutrinos, Geoneutrinos, Low Energy Neutrinos, MSW-LMA, SSM}

 
\maketitle

\section{Solar Neutrinos}
\label{sec:solar}

\begin{figure}[th]
\includegraphics[width=.47\textwidth]{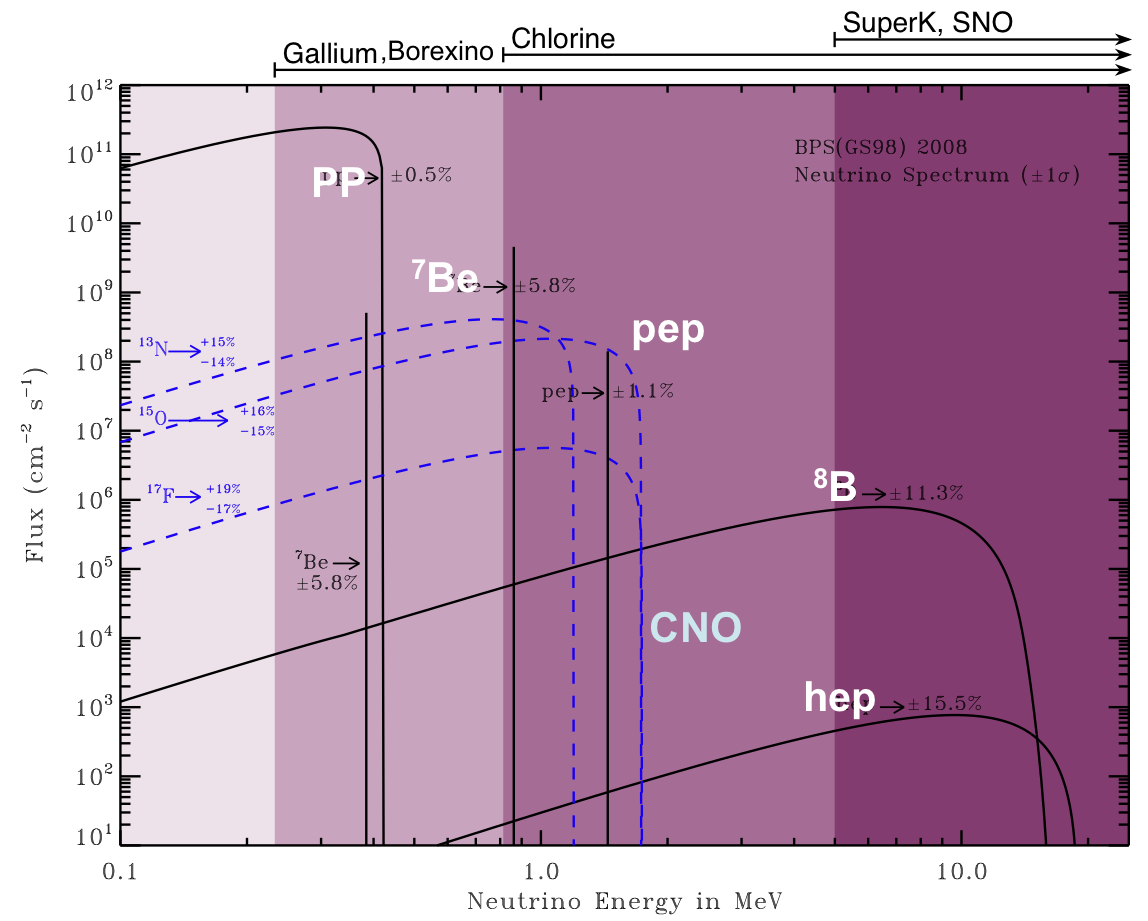}
\includegraphics[width=.51\textwidth]{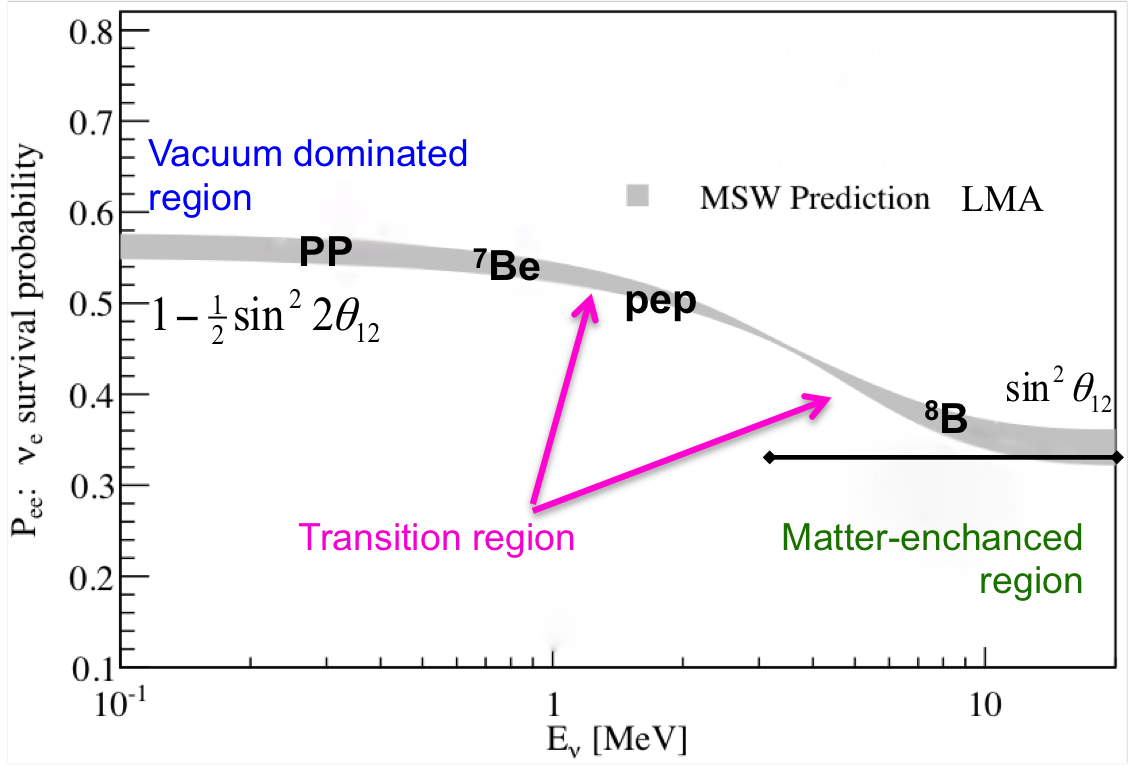}
\caption{Left: solar neutrino spectrum \cite{SHP11}. Right: survival probability for solar electron neutrinos $P_{\textrm{ee}}$ as predicted by the MSW-LMA model.}
\label{fig:ssm_lma}
\end{figure}

Solar neutrino physics lies at the crossing between stellar astrophysics and particle physics beyond the Standard Model of Electroweak interactions. 
The Standard Solar Model describes the thermonuclear processes in the core of the Sun and foresees the emission of neutrinos from different species belonging to two groups of reactions, the so called pp-chain and CNO cycle. 
The expected spectrum can be seen in Figure~\ref{fig:ssm_lma} (left). 

The long standing Solar Neutrino Problem, consisting of reduced neutrino fluxes detected on Earth by many experiments in comparison to the model previsions, was solved during the last decade with the establishment of the neutrino oscillation mechanism. 
Neutrinos, which are no longer believed to be massless, are emitted by the Sun as electron flavor leptons but propagate as mass eigenstates and arrive on Earth as a combination of different flavor eigenstates, to which detectors are differently sensitive. 
The oscillation is enhanced by resonance in the solar matter known as the MSW effect\cite{msw}.
A global analysis of solar and reactor neutrino observations indicate a best fit point in the parameter space of the oscillation $(\Delta m^{2} = 7.59\pm0.21 \cdot 10^{-5} \textrm{eV}^{2}, \sin^{2}(2\theta_{12}) = 0.861^{+0.026}_{-0.022})$ \cite{pdg10} which is know as the Large Mixing Angle (LMA) solution . 
The survival probability for electron flavor solar neutrinos as a function of energy as predicted in the MSW-LMA model is shown in Figure~\ref{fig:ssm_lma} (right).  

Actually solar neutrino oscillation fall into a wider $3\times3$ oscillation scenario together with the atmospheric neutrino oscillation occurring between mass eigenstates 2-3. 
The complete parametrization is give by the PSMN mixing matrix, analogous to the quark mixing matrix \cite{pdg10}. 
The decoupling into two two-neutrino scenarios relies on the third mixing angle $\theta_{13}$ being null. 
This last hypothesis is being highly investigated in these years as moderate indications of a non zero values have recently appeared and would open up for the observation of CP violation in the leptonic sector \cite{sch11}.

\begin{figure}[th]
\begin{centering}
\includegraphics[width=.5\textwidth]{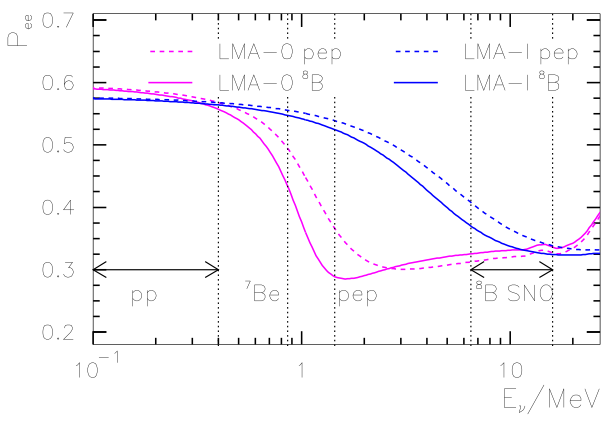}
\caption{The Non Standard Interaction (NSI), labelled as LMA-0 predicts a different $P_{\textrm{ee}}$ in the transition region with respect to LMA \cite{nsi04}}
\label{fig:nsi}
\end{centering}
\end{figure} 

\subsection{Open question in particle physics}
\label{sec:open_particle}

Neutrino physics has now entered the phase in which the precise determination of all oscillation parameters is crucial to unveil new physics. 
In the case of solar neutrinos, the LMA solution has been tested so far in the low energy region (pp and $^{7}$Be), the so called vacuum-dominated regime with errors at the level of 10\% or larger and in the high energy region ($^{8}$B beyond 4-5 MeV), the matter-enhanced region; the intermediate transition region remains basically unexplored.
A number of alternatives to MSW-LMA have been postulated over years, the most popular one being Non Standard neutrino Interactions (NSI) \cite{nsi04}. 
The different survival probability shape can be observed in Figure~\ref{fig:nsi}.
From the experimental point of view there are multiple ways to shed light among the different models. 
The first step is reducing the error on the monochromatic emission of $^{7}$Be at 862keV which would also indirectly constrain the pp flux as this is so far only know from gallium radiochemical experiments\footnote{These experiments had no spectroscopical capabilities. Instead their signal was composed by undistinguished interactions of all neutrinos above their detection threshold of 233keV. Accurately subtracting the $^7$Be and other species leaves the pp flux emerge.} 
The detection of high energy $^8$B neutrinos started with Cherenkov detectors with a threshold of 7.5MeV, which has recently been reduced to 3.5MeV. 
Now scintillator detectors have lowered this threshold to 3MeV and are trying to go down to 2MeV with the goal of observing the upturn of the survival probability foreseen by the LMA solution (Figure~\ref{fig:ssm_lma}, right). 
Finally a measurement of the pep monochromatic emission at 1.2MeV would be a direct test of the transition region.

\subsection{Open question in solar physics}
\label{sec:open_solar}

At the same time, as the neutrino nature is becoming more and more known, neutrinos can play again the role of probes of the Sun interior. 
In particular the fundamental pp-neutrinos are bound to the Sun light emission by the luminosity constraint \cite{SHP11}. 
However the light takes $\sim 10^{5}$y to reach us, while neutrinos escape from the solar matter undisturbed.
Consequently a precision measurement of pp-neutrinos would test the stability of the Sun over this long time scale.
The SSM loosely predicts the contribution of the CNO cycle to the energy balance of the star to be $<1\%$ \cite{SHP11}. 
Detection of the so far unseen CNO neutrinos is therefore a major goal with important cosmological implications.
Another open problem in solar modeling is the so called "Solar Metallicity Puzzle". 
Abundance of isotopes such as C, N, O, Ne and Ar is modeled starting from surface abundance measurements. 
The "old" results, dated 1998 \cite{gs98} and based on 1D modelling, have been improved with the help of 3D modelling (based on the same set of measurements) in 2005 and reviewed in 2009 \cite{ags09} indicating a lower metallicity of the Sun. 
However while the High-metallicity model is perfectly able to reproduce the sound speed profile in the Sun as determined by helio-sysmological measurements, the Low-Metallicity model fails to match this constraint \cite{ser09lowz}. 
Models with different metallicities foresee different neutrino emissions, calling neutrino detectors to solve this ambiguity.

\section{Recent results from solar neutrino experiments}
\label{sec:results}

\begin{figure}[th]
\begin{minipage}{0.25\textwidth}
\includegraphics[width=\textwidth]{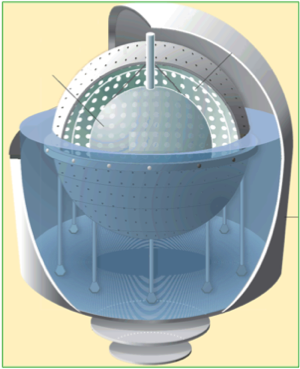}
\end{minipage}
\hfill
\begin{minipage}{0.225\textwidth}
\includegraphics[width=\textwidth]{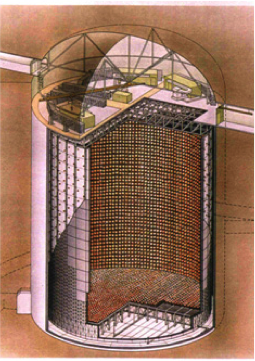}
\end{minipage}
\hfill
\begin{minipage}{0.23\textwidth}
\includegraphics[width=\textwidth]{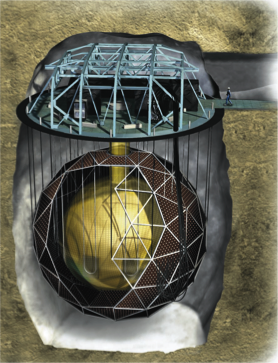}
\end{minipage}
\hfill
\begin{minipage}{0.22\textwidth}
\includegraphics[width=\textwidth]{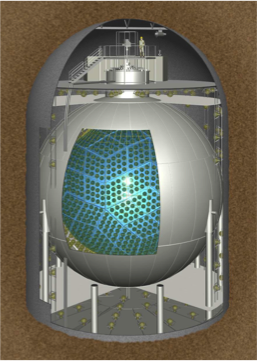}
\end{minipage}
\caption{Sketches of principal solar neutrino detectors. From left to right: Borexino \cite{bx08det}, Super-Kamiokande \cite{sk03det}, SNO \cite{sno99det}, KamLAND \cite{kl10cosm}. }
\label{fig:detectors}
\end{figure}

There are four major experiments (Figure~\ref{fig:detectors}) contributing to results nowadays and many being constructed or proposed.
They rely on two complementary techniques: Water Cherenkov for Super-Kamiokande and SNO liquid scintillator for Borexino and KamLAND.

\subsection{Borexino}
\label{sec:bx}

Borexino is an organic liquid scintillator detector located in the underground Gran Sasso National Laboratory (LNGS) in central Italy under a limestone coverage of $\sim$ 1300 m ($\sim$ 3800\,m\,w.e). 
Data taking started in 2007 and led to the first observation of the $^{7}$Be monochromatic line.

\subsubsection{The detector}
\label{sec:bx_detector}

The Borexino detector \cite{bx08det} was designed to have very low intrinsic background.  
The active target for this measurement is the Inner Detector (ID).
It's central scintillation volume consists of 278\,t
of liquid scintillator, contained in a spherical transparent 8.5\,m diameter nylon Inner Vessel (IV). 
The surrounding 13.7\,m diameter Stainless Steel Sphere (SSS) 
holds 2212 inward-facing 8'' PhotoMultiplier Tubes (PMT) for a 30\% optical coverage.
The ID is surrounded by a powerful muon Outer Detector (OD) \cite{bx11mv},
a high domed steel tank filled with 2\,100\,t of ultra-pure water and instrumented with 208 PMTs which detect the muon Cherenkov emission. 

\begin{figure}[th]
\includegraphics[width=.49\textwidth]{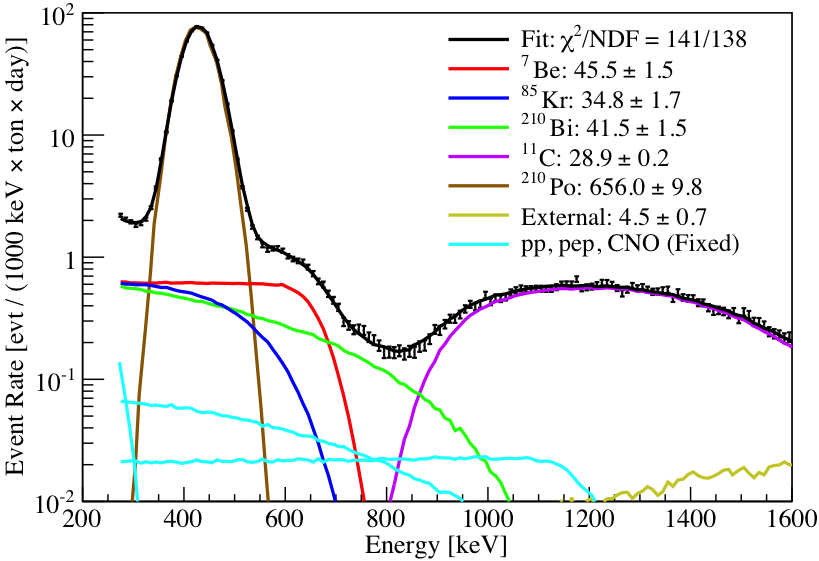}
\includegraphics[width=.49\textwidth]{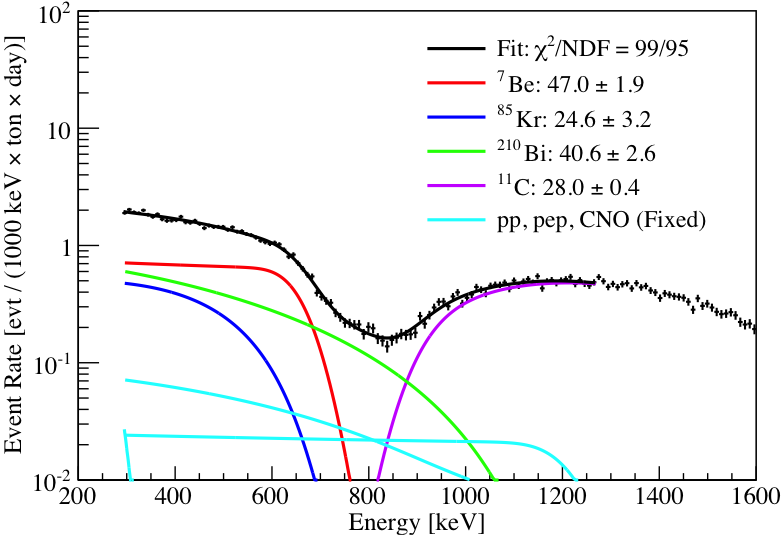}
\caption{Borexino energy spectrum with MC (left) and analytic (right) fits for $^7$Be analysis \cite{bx11be}.}
\label{fig:bx_7be_fit}
\end{figure}

\subsubsection{Flux of 7Be neutrinos}
\label{sec:bx_be}

The radioactive decays of remaining isotopes in the scintillator and in the surrounding detector's material cannot be disentangled on an event-by-event basis from neutrino interactions. 
Consequently the neutrino flux is extracted from a fit to the energy spectrum, which accounts for all signal and background components (see Figure~\ref{fig:bx_7be_fit}). 
In order to limit the impact of external background, a fiducial volume is defined inside the IV via the reconstruction of the position of the events by fitting the arrival photon time distribution on the PMTs.

The first measurement of $^7$Be neutrino flux came after only three months of data and was revised a year later with an error of ~10\% \cite{bx08be}. 
To reduce the systematic error, during the past two years the collaboration performed a thorough set of calibration campaigns using $\alpha$, $\beta$, $\gamma$ and neutron sources deployed in different locations within the IV.
The precise energy calibration of the detector up to about 2.5MeV reduced the energy scale uncertainty to better then 1.5\%; 
while the calibration of the position reconstruction algorithm allowed a determination of the fiducial volume to -1.3\%/+0.5\%
The spectral fit after 741 live days \cite{bx11be} is performed in two ways, both fixing pp, pep, CNO and $^8$B to the signals foreseen by the SSM (SP11 with High-Z), leaving free the $^7$Be signal and backgrounds from $^{85}$Kr($\beta^{-}$), $^{210}$Bi($\beta^{-}$), $^{11}$C($\beta^{+}$), $^{210}$Po($\alpha$). 
The first approach uses MonteCarlo generated curves in the [250-1600]keV range and accounts for external $\gamma$'s from the PMTs in the rightmost region, while the second approach uses analytical shapes in the [300-1250]keV range obtained after statistical subtraction of the $\alpha$ events ($^{210}$Po) as identified by pulse shape analysis. 
The two methods yield fully consistent results and the difference in the flux value returned has been included in the systematics which now accounts to $\sim$3.5\%.
The final result is 46.0$\pm$1.5(stat)$^{+1.5}_{-1.6}$(syst) counts per day in 100t, for the first time below 5\% accuracy and for the first time below the theoretical error of $\sim$ 7\%. 
Looking directly at the $^7$Be neutrino flux in units of $10^9$ cm$^{-2}$s$^{-1}$ the Borexino measured flux is 3.10$\pm$015 or 4.84$\pm$0.24 after correcting for neutrino oscillation with the best-fit parameters from global analysis (\cite{pdg10}). 
The SSM-High Z (Low-Z) model foresees instead 5.00$\pm$0.35(4.56$\pm$0.32). 
Considering the High-Z model as reference the measured to expected flux ratio is therefore 0.62$\pm$0.05 corresponding to a survival probability of 0.5$\pm$0.07, while including oscillations the ratio becomes $f(^7\textrm{Be})=0.97\pm0.09$.

\begin{figure}[th]
\begin{centering}
\includegraphics[width=.5\textwidth]{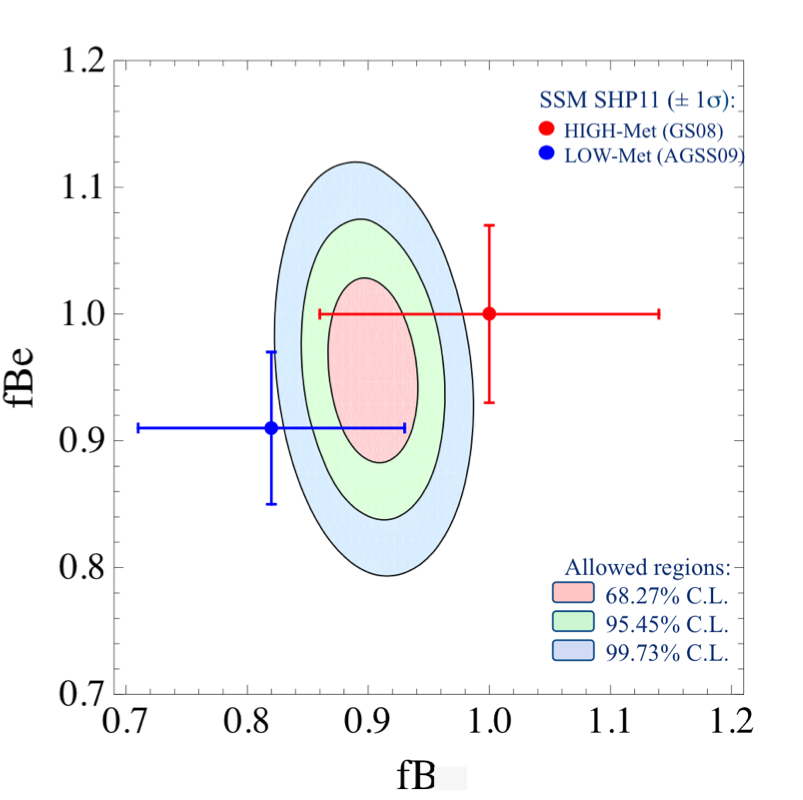}
\caption{$^7$Be and $^8$B measured values as a fraction of High-Z SSM prediction, compared to the two metallicity expectations \cite{ca11met}.}
\label{fig:metallicity}
\end{centering}
\end{figure}

Figure~\ref{fig:metallicity} shows the combined confidence level contours for $^7$Be and $^8$B measurements including the latest Borexino measurements expressed as a fraction of the reference High-Z model. 
Predictions from the two models with theoretical error bars are also shown. 
As it can be seen the present data do not allow to discriminate between the two models.

\begin{figure}[th]
\includegraphics[width=.49\textwidth]{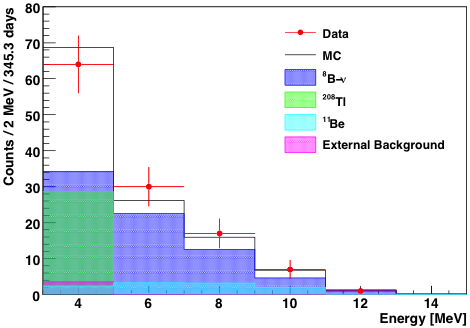}
\hfill
\includegraphics[width=.50\textwidth]{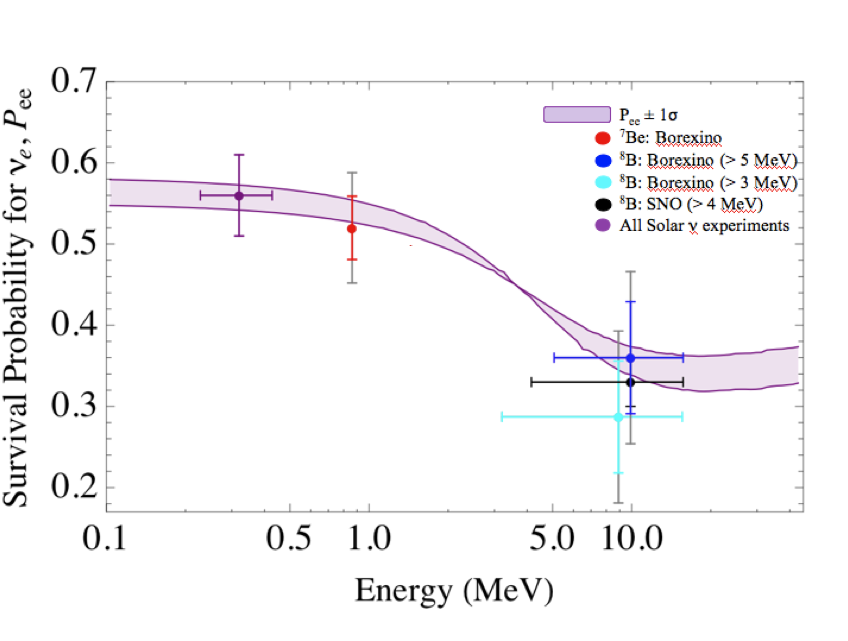}
\caption{Left: Borexino $^8$B energy spectrum. Right: available measurements of $P_{\textrm{ee}}$ after Borexino \cite{bx10B}.}
\label{fig:bx_8b_lma}
\end{figure}

Within a global analysis the Borexino results on $^7$Be flux give a strong contribution in constraining the fundamental pp neutrino flux and reducing its error: $\Phi_{\textrm{pp}} = 6.06^{+0.02}_{-0.06} \cdot 10^{10}$ cm$^{-2}$s$^{-1}$ ($f(\textrm{pp}) = 1.013^{+0.003}_{-0.010}$). 
Figure~\ref{fig:bx_8b_lma} (right) shows the impact of the Borexino measurement on the validation of the MSW-LMA model in the vacuum-dominated regime. 
In addition the CNO neutrino flux can be limited to $\Phi_{\textrm{CNO}} < 1.3 \cdot 10^{9} \textrm{cm}^{-2}\textrm{s}^{-1}$ ($f(\textrm{CNO}) <2.5$) at 95\% C.L. 
Consequently the contribution to the SSM luminosity from CNO cycle can be limited to $<1.7\%$ at 95\% C.L. 

\begin{figure}[th]
\includegraphics[width=.46\textwidth]{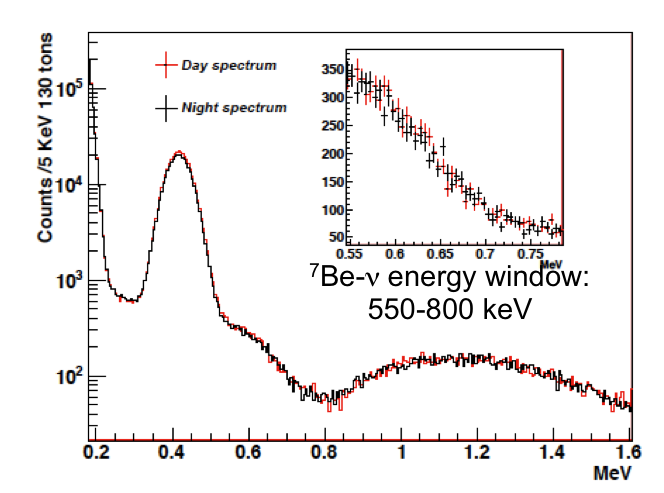}
\hfill
\includegraphics[width=.53\textwidth]{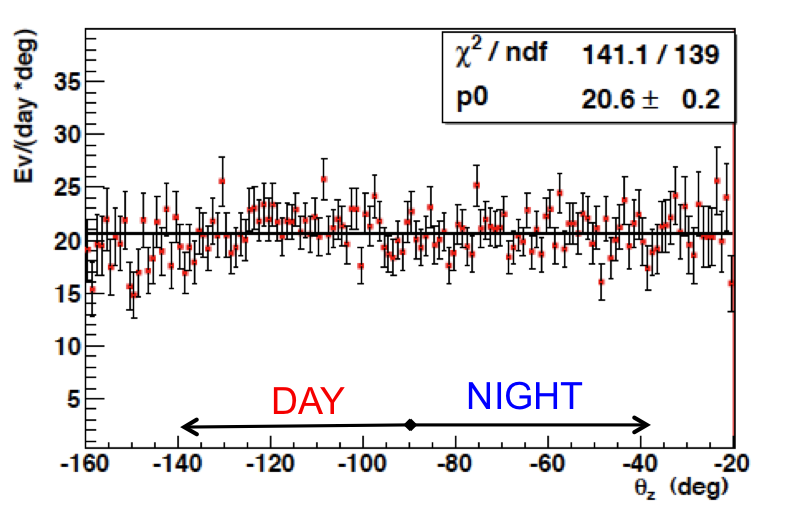}
\caption{Borexino day and night energy spectra and nadir angle distribution for the $^7$Be energy region \cite{bx11dn}.}
\label{fig:bx_dn}
\end{figure}

\subsubsection{Day-Night asymmetry}
\label{sec:bx_dn}

The Borexino data set has been inspected also for a possible day-night effect. 
Figure~\ref{fig:bx_dn} shows the day and night Borexino spectra, after normalizing for the exposure (385.5 live daytime and 363.6 live nighttime).
The energy region considered for the extraction of the day-night parameter is the $^7$Be energy window (550-800keV). 
The nadir angle distribution for events in this range is also shown where the exposure function used for normalization also accounts for the expected 3\% seasonal modulation due to the earth orbit eccentricity, which could otherwise introduce a fake day-night effect. 

\begin{figure}[th]
\includegraphics[width=.59\textwidth]{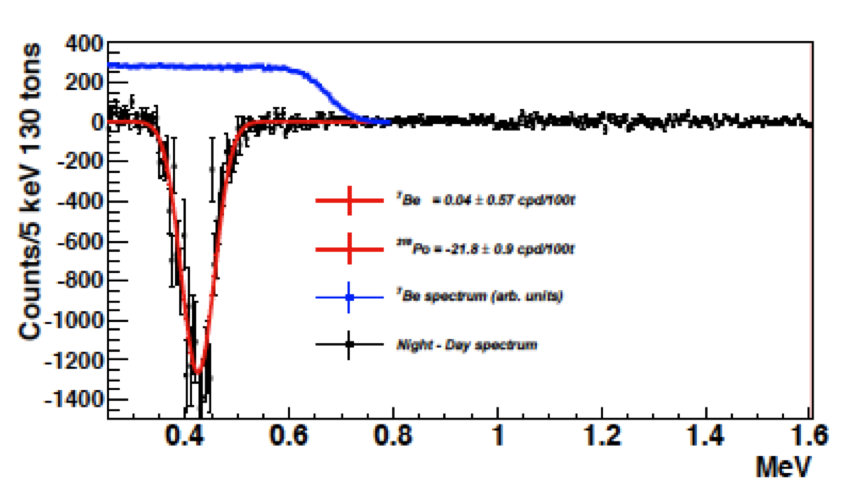}
\includegraphics[width=.39\textwidth]{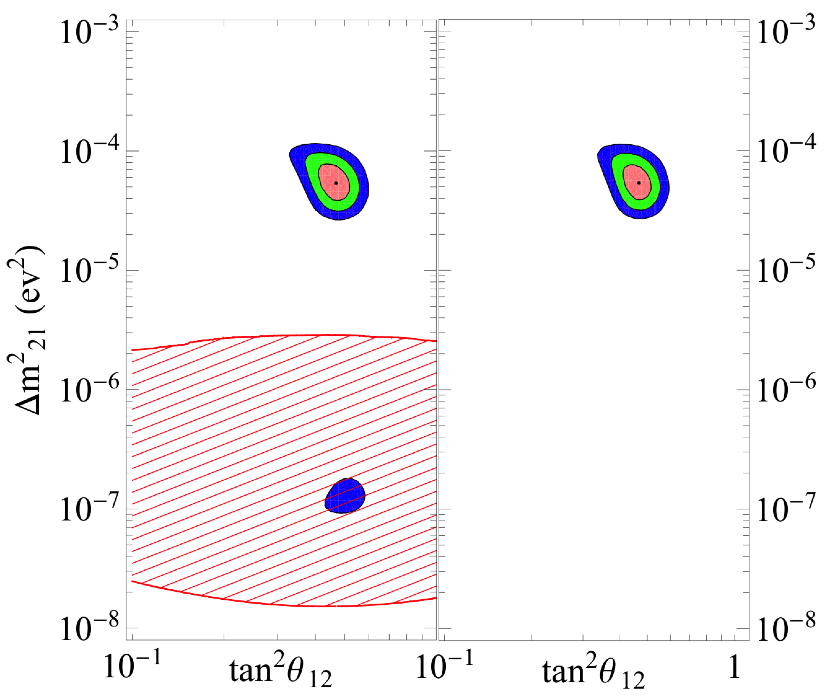}
\caption{Left: Borexino night-day subtracted spectrum. Right: the Borexino non observation of  $A_{\textrm{dn}}$ in the global analysis allows to reject the LOW solution at $\Delta m_{12}^2 \sim 10^{-7} \textrm{eV}^2$ \cite{bx11dn}.}
\label{fig:bx_dn_sub_excl}
\end{figure}

The Day-Night asymmetry parameter is defined as $A_{\textrm{dn}} = 2 \frac{R_{\textrm{n}} - R_{\textrm{d}}}{R_{\textrm{n}}+R_{\textrm{d}}} = R_{\textrm{diff}}/R$  where $R_{\textrm{n}}$ and $R_{\textrm{d}}$ are the $^7$Be night and day fluxes respectively. 
The most effective way of extracting this parameter is to subtract the day spectrum from the night one (after normalization) and fit the resulting spectrum searching for a contribution with the $^7$Be Compton-like shape.
Figure~\ref{fig:bx_dn_sub_excl} (left) shows this procedure. The negative peak around 400keV is not surprising: the $^{210}$Po alpha contribution is more present in the day spectrum then in the night one as the decay time is $\sim$200d and data taking started at the beginning of a summer. 
No $^{7}$Be is found in the subtracted spectrum and the resulting $A_{\textrm{dn}}=0.001\pm0.012(\textrm{stat})\pm0.007(\textrm{syst})$.
The systematic error is given by the variation of the $^{210}$Bi contamination through the dataset and by the result variation in applying the fit procedure after $\alpha$ statistical subtraction (as described in section~\ref{sec:bx_be}).

Figure~\ref{fig:bx_dn_sub_excl} (right) shows the impact of this measurement. 
When neutrino oscillations with MSW effect started to be a concrete solution to the solar neutrino problem, four minima were identified by the global analyses fits in the $\Delta m^2_{12}, \tan^2(\theta_{12})$ parameter plane, one of which is LMA. 
Among the the others, the so called LOW solution foresees a pronounced $A_{dn}$ due to electron neutrino regeneration effects in the earth matter at night. 
Ruling out this solution so far has required the use of KamLAND antineutrino data (see section~\ref{sec:kl_reactors}), while it remained allowed in solar-only global analyses although at $\Delta \chi^2 = 11.8$.  
Thanks to the Borexino result the LOW solution is now ruled out at $>8.5\sigma$ without the need to assume CPT invariance and the LMA best fit point is ($\Delta^{2}_{12}=5.2\cdot10^{-5} eV^{2}, \tan^{2}\theta_{12} = 0.46$).
In addition the hypothesis of mass-varying neutrinos \cite{hol09}, also implying day night asymmetry is ruled out at $\sim 10\sigma$.

\subsubsection{Flux of 8B neutrinos}
\label{bx_b}

The Borexino collaboration has also performed a measurement of the $^8$B neutrino flux and spectrum (Figure~\ref{fig:bx_8b_lma}, left). 
Although Borexino is small in comparison to Water Cherenkov detectors for a statistically competitive measurement, the threshold reached by Borexino is the lowest ever achieved: 3MeV for the electron recoil energy.
The rate measured is 0.22$\pm$0.04(stat)$\pm$0.01(syst) c/d/100t corresponding to a flux of $\phi^{\textrm{ES}}_{\textrm{exp}} = 2.4\pm0.4\pm0.1 \cdot 10^6 \textrm{cm}^{-2}\textrm{s}^{-1}$ or $\Phi^{\textrm{ES}}_{\textrm{exp}}/\Phi^{\textrm{ES}}_{\textrm{th}} = 0.88\pm0.19$ with respect to High-Z SSM (\cite{ser09lowz}). 
The relevant contributions to the systematic error are: energy threshold definition ($^{+3.6\%}_{-3.2\%}$), fiducial mass ($\pm{3.8\%}$) and energy resolution ($^{0.0\%}_{-2.5\%}$). 
The analysis has been performed also setting the energy at 5MeV, attesting the full compatibility with the results of Water Cherenkov detectors described below.
The importance of performing this measurement with Borexino can be seen in Figure~\ref{fig:bx_8b_lma} (right).
Borexino has measured $P^{\textrm{vac}}_{\textrm{ee}}(^7\textrm{Be})=0.52^{+0.07}_{-0.06}$ and $P^{\textrm{matt}}_{\textrm{ee}}(^8\textrm{B})=0.29\pm0.10$ with $P^{\textrm{vac}}_{\textrm{ee}}-P^{\textrm{matt}}_{\textrm{ee}}=0.23\pm0.07$. 
For the first time the same detector has measured the survival probability of electron neutrinos from the sun in the two energy regions foreseen by the MSW-LMA, the vacuum-dominated and the matter-enhanced regions, and obtaining different results, a striking confirmation of the model.

\subsection{Super-Kamiokande}
\label{sec:sk}

The Japanese Super-Kamiokande (SK) is a Water Cherenkov detector located 1000m (2700m w.e.) underground and since fifteen years plays a fundamental role in most neutrino physics searches including solar, atmospheric, supernova and long baseline neutrino studies. 
Concerning solar neutrinos the detector focuses on the detection of $^8$B neutrinos via elastic scattering.

\subsubsection{The detector}
\label{sec:sk_detector}

The SK \cite{sk03det} Inner Detector is a 50kton water tank with cylindrical geometry and equipped with 11129 20'-PMTs.  
A fiducial volume of 22.5kton is standard for solar neutrino studies. 
An outer layer of water acts as muon detector being equipped with additional 1885 8'-PMTs.

SK started data taking in 1996 and produced brilliant results in the following years (\cite{sk06I}). 
The energy threshold was 5MeV for solar neutrinos. 
An accident occurred in 2001 during a maintenance operation breaking about half of the PMTs. 
A temporary situation named SK-II had the remaining PMTs rearranged to cover the full surface and operated with a 7MeV threshold for $^8$B neutrinos in years 2003-2005. 
In 2006 SK-III restarted with the full coverage and threshold of SK-I .
Since 2008 the detector is in SK-IV phase with upgraded electronics and water plants.
The current threshold is at 4.5MeV and is targeting 4MeV for the nearest future.
 
\begin{figure}[th]
\includegraphics[width=.46\textwidth]{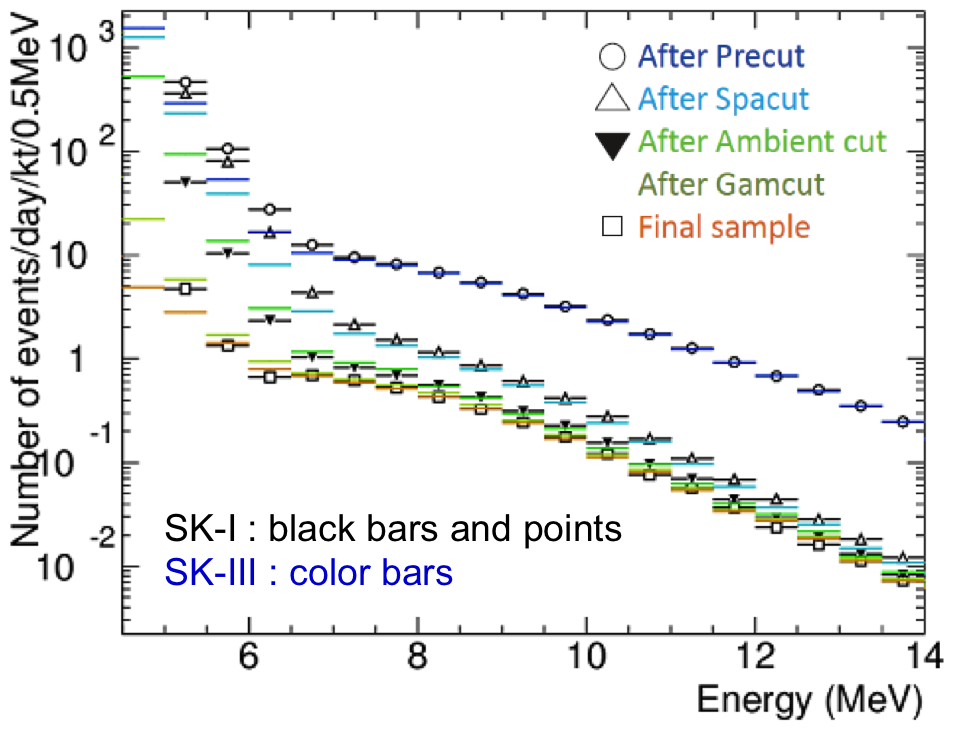}
\hfill
\includegraphics[width=.53\textwidth]{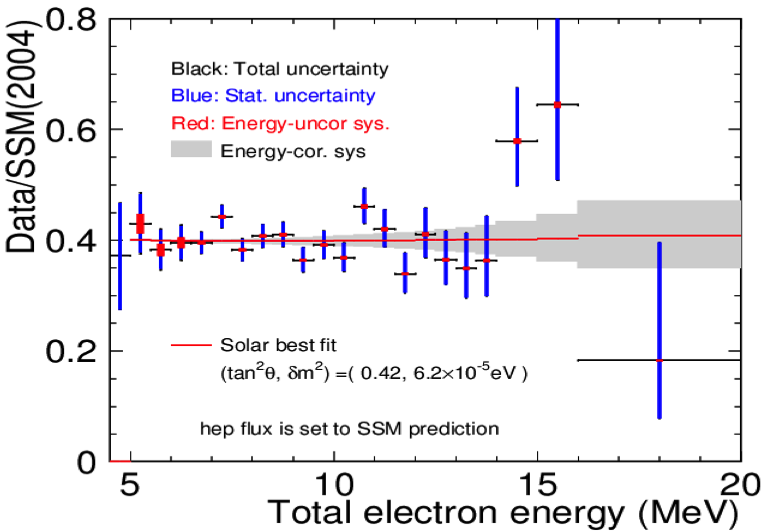}
\caption{SK-III energy spectrum \cite{sk10III}. Left: effect of purification in comparing with SK-I spectrum. Right: after normalized to SSM predictions.}
\label{fig:skIII}
\end{figure}

\subsubsection{Results from SK-III phase}
\label{sec:sk_III}

The SK-III phase \cite{sk10III} featured improved water circulation and purification system resulting in reduced Radon in the FV compared to SK-I phase. 
The effect can be seen in the lowest energy bin [5.0,5.5]MeV in Figure~\ref{fig:skIII}.
An additional bin is available in the spectrum at [4.5,5.0]MeV, where solar neutrino signal is present with $4\sigma$ significance. 
However the systematics for this bin is still being evaluated and is therefore not included in the fit. 
The SK-III spectrum shows no distortion from the shape predicted by the SSM (Figure~\ref{fig:skIII}).
The $^8$B neutrino flux is $2.32\pm0.04(\textrm{stat})\pm0.05(\textrm{syst})\cdot10^6\textrm{cm}^{-2}\textrm{s}^{-1}$ to be compared with the SK-I result (\cite{sk06I}) $2.38\pm0.02(\textrm{stat})\pm0.08(\textrm{syst})\cdot10^6\textrm{cm}^{-2}\textrm{s}^{-1}$. 
As it can be seen the results are fully consistent and SK-III although limited by a larger statistical error it features a significantly reduced systematic error thanks to the improved timing calibration of the detector.

\begin{figure}[th]
\begin{centering}
\includegraphics[width=.49\textwidth]{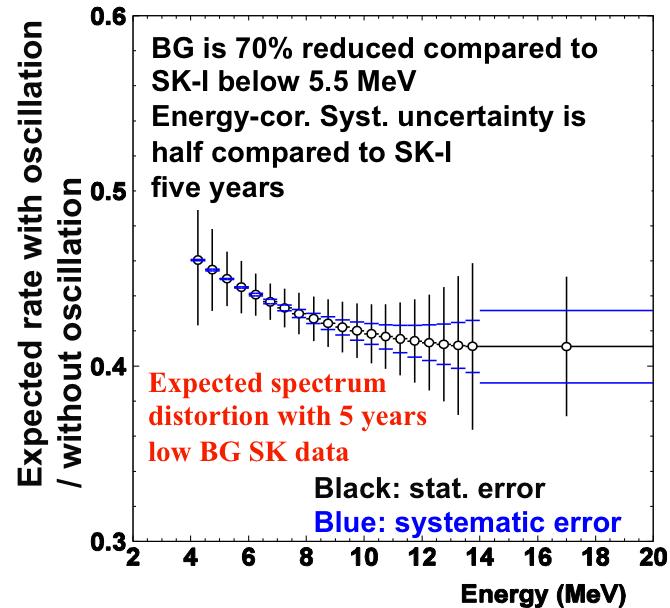}
\caption{Expected spectrum for SK-IV phase after 5y of data taking \cite{sk10IV}.}
\label{fig:skIV}
\end{centering}
\end{figure} 

\subsubsection{Prospects for the SK-IV phase}
\label{sec:sk_IV}

At the end of the SK-III phase, the electronics of has been upgraded achieving a wider dynamic range ad the DAQ has been modified in view of a transition from a hardware trigger system to a software one with lower threshold possibilities. 
In addition the water plants have been modified for a better control of the temperature of the water at the inlet of the detector's tank and a consequent further reduction of Radon within the fiducial volume.
All these improvements aim to achieve a 4MeV threshold with half the systemic error of SK-I 
and test the upturn of the $^8$B neutrino spectrum within 5 years of data taking. 
The expected spectrum can be seen in Figure~\ref{fig:skIV} \cite{sk10IV}.

\subsection{SNO}
\label{sec:sno}

The Sudbury Neutrino Observatory (SNO) is a 1\,kton heavy water Cherenkov detector located in the Sudbury mine (Canada) at the formidable depth of 2092m ($\sim$6010m w.e.)  which has taken data between 1999 and 2006 giving fundamental contributions to solar neutrino physics.

\subsubsection{The detector}
\label{sec:sno_detector}

The detector \cite{sno99det} is developed for $^8$B solar neutrino detection and has the advantage of multiple interaction channels. 
Aside from neutrino-electron elastic scattering, which is diversely sensitive to neutrino flavors, heavy water allows 
for charged current ($\nu_e+D\rightarrow e^-+p+p$) and neutral current ($\nu_x+D\rightarrow\nu_x+p+n$) interactions, 
which are sensitive to electrons neutrinos only and to all neutrinos regardless of flavor, respectively.
The three signals are extracted by statistical fit of the energy spectrum. 
In 2002 measured different fluxes in the neutral and charged current interaction channels: the proof of a flavor changing mechanism operating for solar neutrinos.

The target mass of 1000t of heavy water is contained in a 12m-diameter acrylic vessel and is watched by 9500 PMTs allocated on a 18m support structure for a total photocathode coverage of $\sim$60\%.
Regular water fills all volumes outside the acrylic vessel and inside the detector's cave, shielding the sensible volume from environmental radioactivity.

The data taking occurred in three successive phases. 
In 2001, after a phase-I with pure D$_2$O, salt was dissolved into the water to enhance the detection of neutrons from neutral current interactions. 
Neutron capture on $^{35}$Cl in fact features a higher cross section and a higher energy (8.6MeV) of the gamma cascade -- standing better out of background -- with respect to the 6.5MeV gamma from n capture on D. 
At the end of 2003, the salt was removed and during 2004 a set of NaI proportional counters have been deployed in strings throughout the active volume. 
This allowed n-capture via $n+^3He\rightarrow t+p$ reaction with a cross section of 5330b and the possibility of event by event separation.
However the phase-III, whose analysis has been finalized this year \cite{sno11ph3}, did not improve significantly the brilliant results of the previous two phases \cite{sno07ph1, sno05ph2}.

\subsection{Low Energy Threshold Analysis}
\label{sec:sno_leta}

The most interesting achievement of the SNO collaboration in last two years is the Low Energy Threshold Analysis (LETA) \cite{sno10leta}, the complete re-analysis of the data from the first two phases lowering the energy threshold of 5.0MeV (phase-I) or 5.5MeV (phase-II) to 3.5MeV with a significant gain in statistics ($\sim +70\%$). 
In addition LETA is also an advanced multivariate analysis that fits the binned distributions of four variable simultaneously: the angle between the reconstructed Cherenkov track and the direction of the Sun ($\cos (\theta_{\textrm{Sun}})$), a shape parameter indicating the sphericity of the PMT hit distributions in the event ($\beta_{14}$), the radial distribution ($R^3$) and of course the reconstructed energy of the event ($T_{\textrm{eff}}$).

\begin{figure}[th]
\includegraphics[width=.49\textwidth]{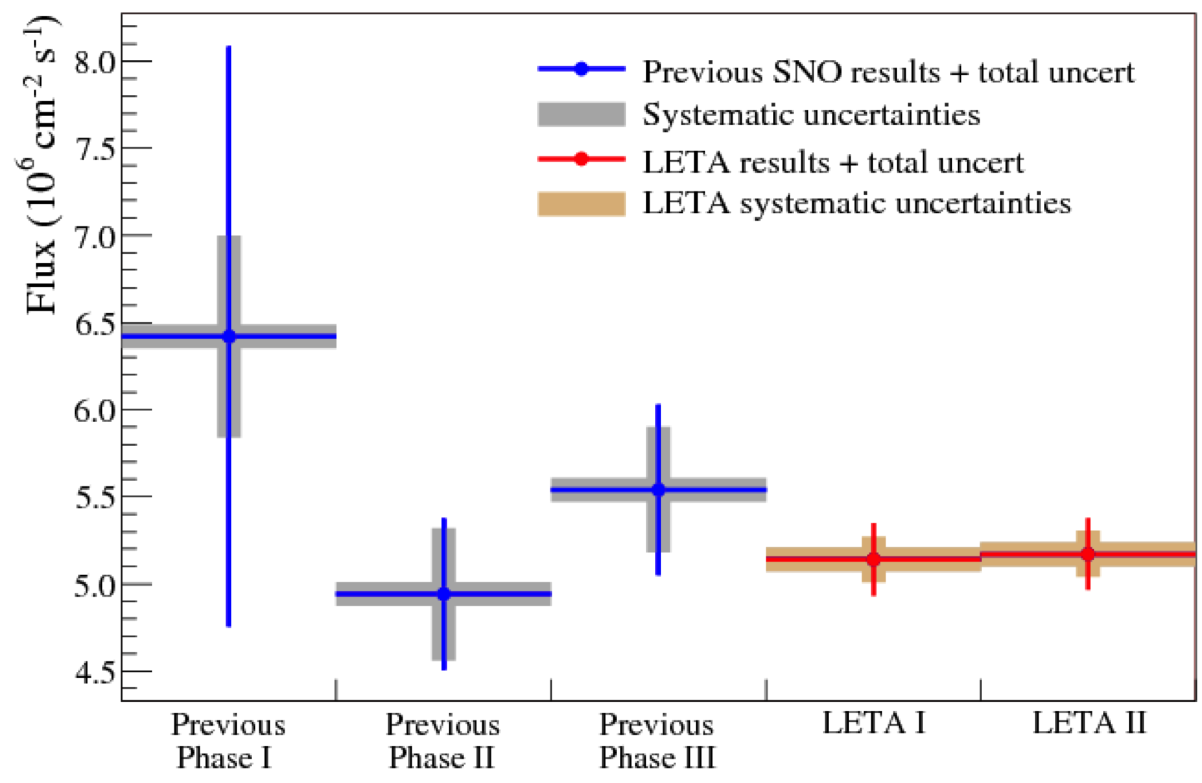}
\includegraphics[width=.49\textwidth]{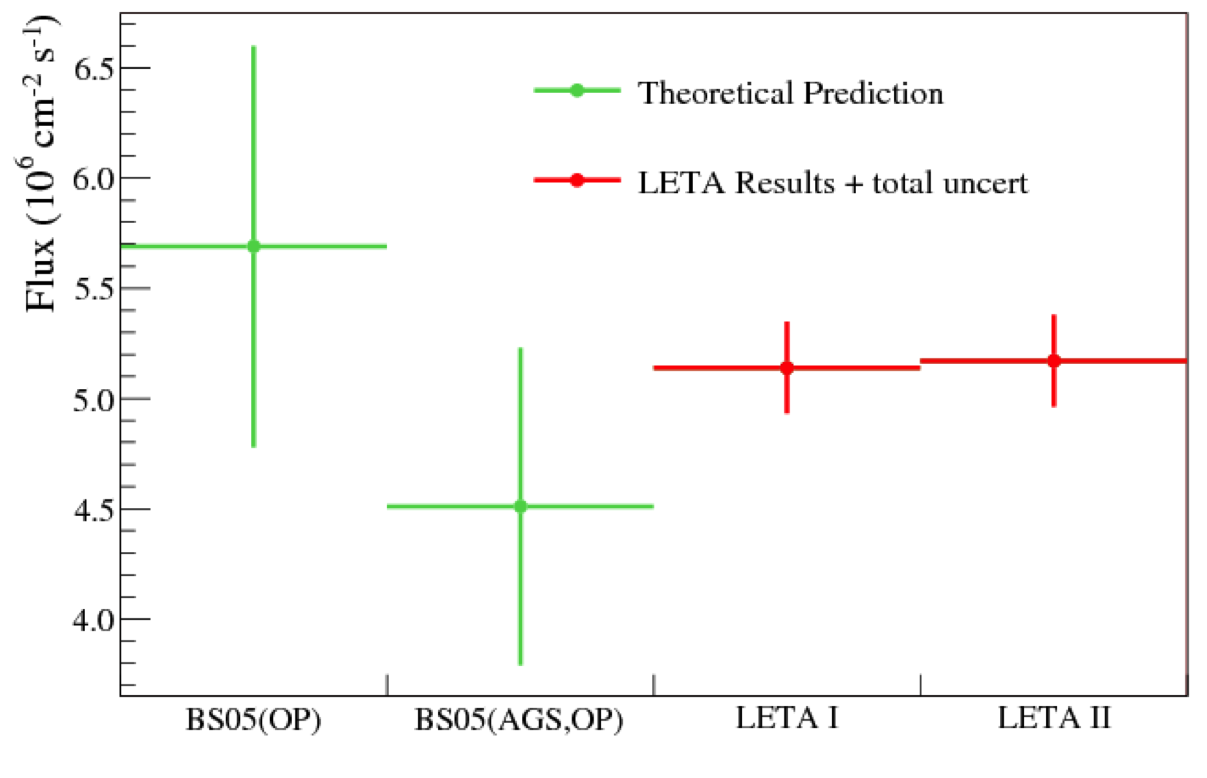}
\caption{SNO LETA results compared to previous analyses of the three phases and to predictions of SSM in the two metallicity cases \cite{sno10leta}.}
\label{fig:sno_leta}
\end{figure}

\begin{figure}[th]
\begin{centering}
\includegraphics[width=.49\textwidth]{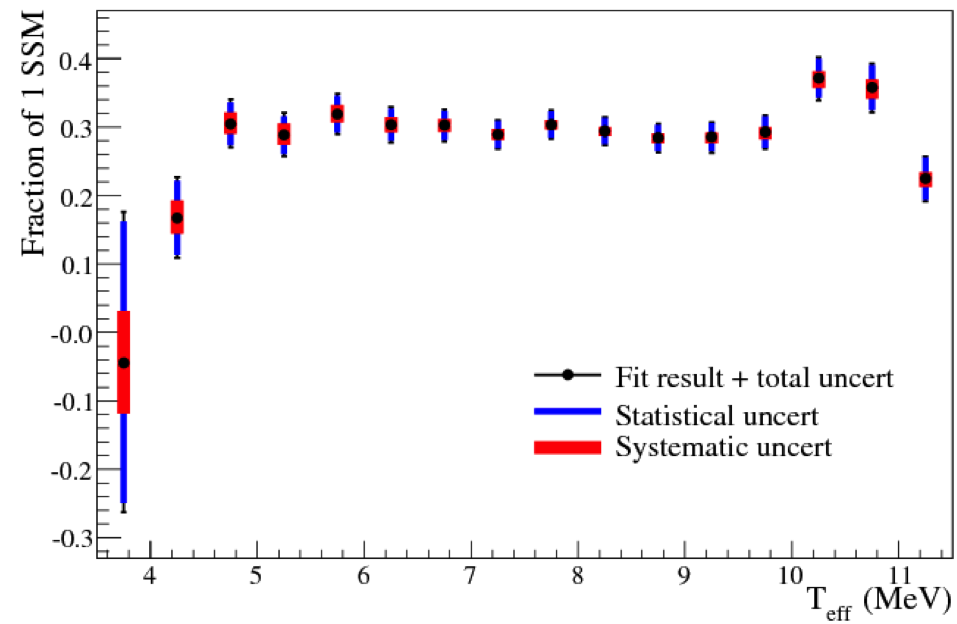}
\caption{SNO LETA Charged Current interaction spectrum \cite{sno10leta}.}
\label{fig:leta_cc}
\end{centering}
\end{figure}

Figure~\ref{fig:sno_leta} shows the striking reduction in errors (both statistical and systematical) achieved with LETA in comparison with the original analyses for the three phases and with the prediction of SSM in the two metallicity cases. 
As noted before (section~\ref{sec:bx_be} and Figure~\ref{fig:metallicity}), 
the measured values lie in between the two prediction and currently allow to disfavor neither of them. 
Finally Figure~\ref{fig:leta_cc} shows the CC spectrum obtained in LETA and, although the small statistics in the lowest energy bin does not allow to draw conclusions, there is no clue of the expected upturn of the $^8$B spectrum, which raises the interest for the upcoming measurement of SNO+ (section~\ref{sec:future_sno+}).

\subsection{KamLAND}
\label{sec:kl}

KamLAND is a liquid scintillator neutrino detector located in the Kamioka mine in Japan and taking data since 2002. 
Its main physics target are electron antineutrinos from nuclear power reactors in Japan and Korea, ranging up to 7MeV. 
The flux-weighted average baseline for KamLAND is $\sim$180km which makes the oscillations directly visible in the spectrum.
Under the assumption of CPT invariance these results can be combined with solar neutrino experiments.
In 2003 KamLAND's spectral distortions led to the identification of oscillations as the flavor-converting mechanism operating on solar neutrinos and of LMA as the best fit solution in the "solar" sector.

\subsubsection{The detector}
\label{sec:kl_detector}

The primary target volume consists of 1 kton of ultra-pure liquid scintillator (LS). 
This inner detector of LS is shielded by a 3.2-kton water-Cherenkov outer detector. 
Scintillation light is viewed by 1325 17-inch and 554 20-inch photomultiplier tubes (PMTs) 
providing 34\% solid-angle coverage. A detailed overview of the detector is given in \cite{kl10cosm}.

\subsubsection{Reactors' antineutrino spectrum}
\label{sec:kl_reactors}

Antineutrino detection in liquid scintillator occurs via the "standard" Reines-Cowan technique: 
the antineutrino interacts via inverse beta decay $\overline{\nu}_e + p \rightarrow n + e^+$ and the signature consists in a prompt signal from positron scattering on top of the annihilation energy and a delayed signal from the neutron capture gamma (2.2MeV on H). 
A typical delay between the two signals in liquid scintillator is of the order of 250\,\textmu s so the coincidence is a powerful handle to suppress background. 
The process has a kinematic threshold of 1.8MeV.
 
\begin{figure}[th]
\includegraphics[width=.42\textwidth]{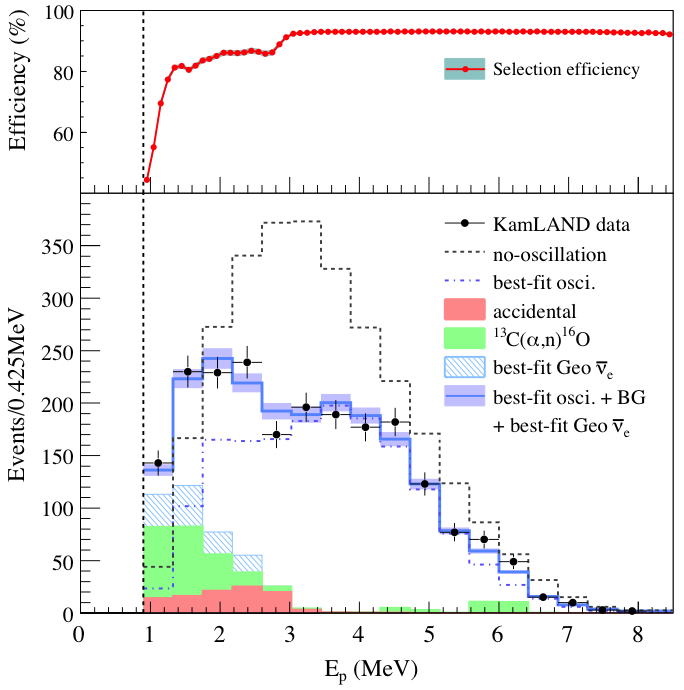}
\includegraphics[width=.57\textwidth]{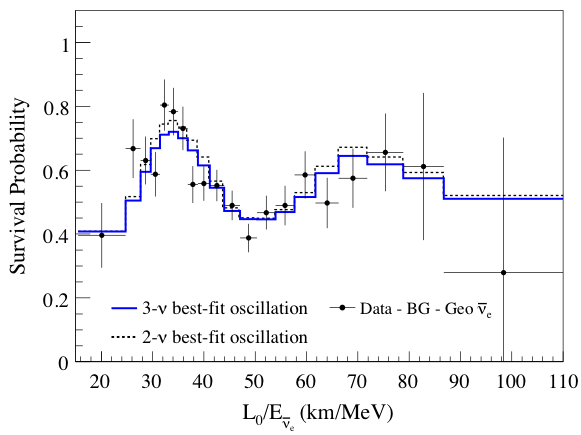}
\caption{KamLAND reactors' data: prompt energy spectrum and L/E distribution \cite{kl11rea}.}
\label{fig:kl_spectrum}
\end{figure}

The most recent update of reactor antineutrino results from the KamLAND collaboration is \cite{kl11rea} and features a data set of 2135\,d live time taken from March 2002 to Apr 2009. 
The spectrum can be seen in Figure~\ref{fig:kl_spectrum}. 
$^{210}$Po is the main radioactive background source as the $\alpha$ particles emitted by its decay are responsible for $^{13}$C($\alpha$,n)$^{16}$O reactions which simulate the coincidence signals of the antineutrino interactions. 
For this reason in 2007 the LS was purified reducing the contamination of $^{210}$Po by a factor of $\sim$20. 
The latest result include $\sim$30\% live time from the post purification period and the background reduction is visible by comparing Figure~\ref{fig:kl_spectrum} to previous editions of the same plot (\cite{kl05rea}). 
Figure~\ref{fig:kl_spectrum} also shows the spectrum as a function of $L_0/E_{\overline{\nu}_e}$ making the oscillations clearly visible.

\begin{figure*}
\begin{center}
\includegraphics[width=\textwidth]{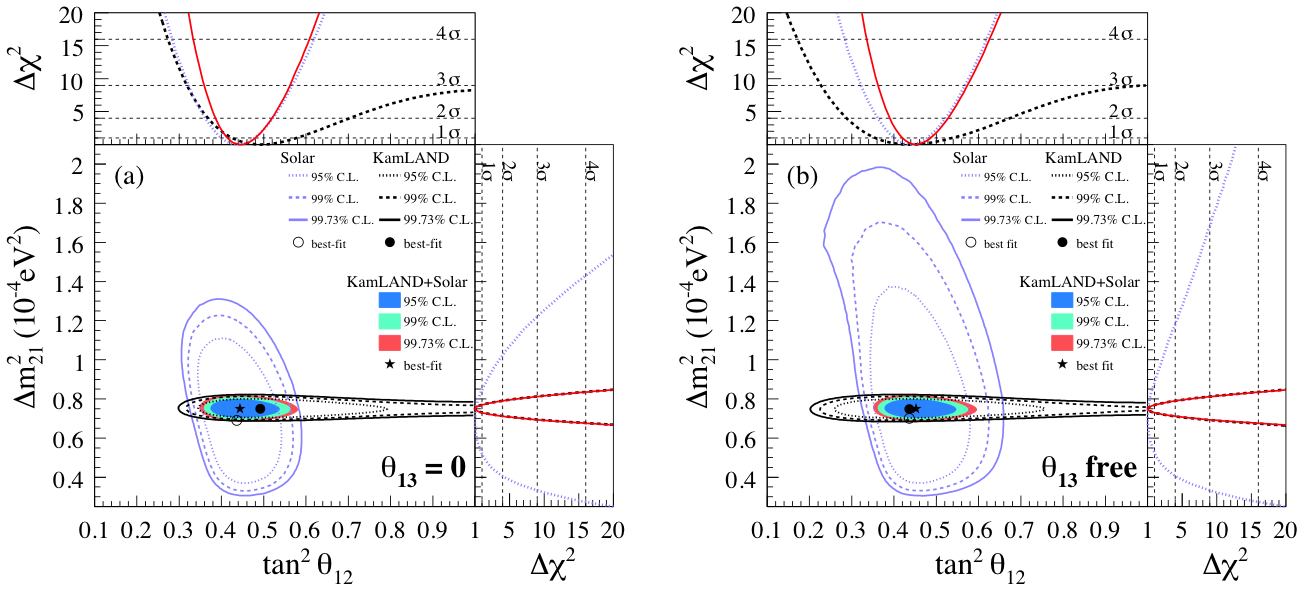}
\caption{$\Delta m^2_{12}, \tan^2 \theta_{12}$ from global analysis using solar data, KamLAND data or combining them all. 2-$\nu$ analysis (left) and 3-$\nu$ analysis (right) are shown \cite{kl11rea}.}
\label{fig:kl_global}
\end{center}
\end{figure*} 

\subsubsection{Global Analysis}
\label{sec:kl_global}

If solar neutrino and reactors' antineutrino data are combined into a global analysis the neutrino oscillation parameters between the first two mass eigenstates can be obtained from the fit \cite{kl11rea}. 
The parameter plane $\Delta m^2_{21}, \tan^2(\theta_{12})$ is shown in Figure~\ref{fig:kl_global} in the case of 2-$\nu$ ($\theta_{13}$ fixed to 0) and 3-$\nu$ ($\theta_{13}$ left free) analysis. 
The best fit points are $\tan^2\theta_{12}= 0.444^{+0.036}_{-0.030}$ and $\tan^2\theta_{12}= 0.452^{+0.035}_{-0.033}$, respectively, while $\Delta m^2_{12} = 7.50\pm0.20 \cdot 10^{-5} \textrm{eV}^2$ in both cases. 
The slight tension existing between the solar-only and the KamLAND-only fit results in the 2-$\nu$ case is released in the 3-$\nu$ case and the fit returns $\sin^2(\theta_{13}=0.02\pm0.016)$. 
This is a modest indication for a non-zero value, but it can be combined with other neutrino physics results obtaining a somewhat stronger significance (see for example \cite{sch11}).

\begin{figure}[th]
\includegraphics[width=.29\textwidth]{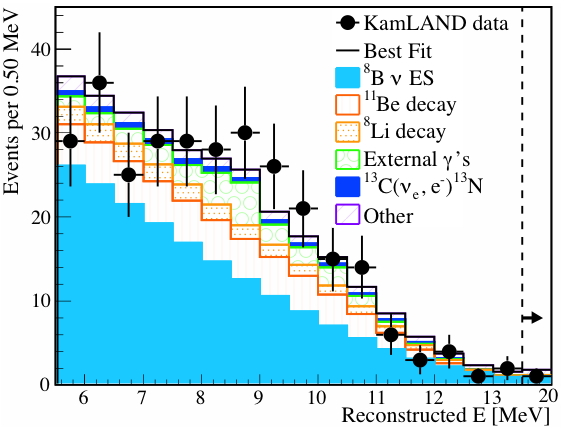}
\includegraphics[width=.70\textwidth]{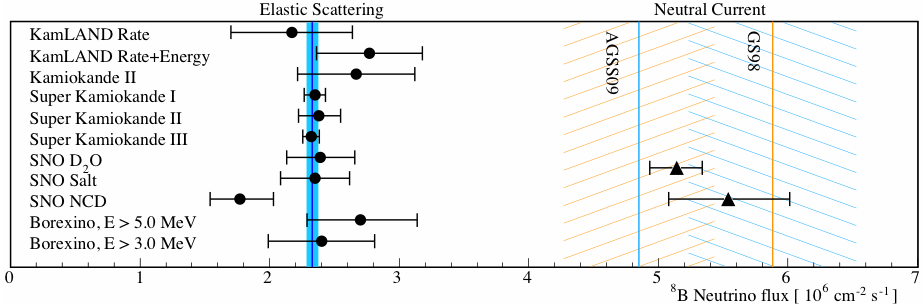}
\caption{Left: KamLAND prompt energy spectrum for $^8$B. Right: comparison all $^8$B flux measurements \cite{kl11B}.}
\label{fig:kl_8b_spectrum_cmp}
\end{figure}

\subsubsection{Flux of 8B solar neutrinos}
\label{sec:kl_b}

KamLAND recently joined the search for solar neutrinos from $^8$B \cite{kl11B} using the data set before the purification of the liquid scintillator, i.e. from Apr. 2002 to Apr. 2007 for a total of 1452d live time. 
This analysis is based on neutrino-electron elastic scattering and does not benefit from the handle of the delayed coincidence with the neutron capture typical of the antineutrino inverse-beta interaction. 
For this analysis it was therefore used a reduced Fiducial Volume of 176.4m$^3$ and a threshold of 5.5MeV. 
The spectrum can be seen in Figure~\ref{fig:kl_8b_spectrum_cmp} (left). 
The extraction of the neutrino flux is attempted with two approaches: a rate only analysis and an unbinned Maximum-Likelihood fit of the energy spectrum. 
The two methods return $\Phi_{\textrm{ES}}^{\textrm{Rate}}=2.17\pm0.26(\textrm{stat})\pm0.39(\textrm{syst})\times10^6\textrm{cm}^{-2}\textrm{s}^{-1}$ and $\Phi_{\textrm{ES}}^{\textrm{Spectrum}}=2.77\pm0.26(\textrm{stat})\pm0.32(\textrm{syst})\times10^6\textrm{cm}^{-2}\textrm{s}^{-1}$, respectively.
The main sources of systematics are the determination of background from cosmogenic radioisotopes such as $^{11}$Be, $^8$Li and $^8$B or from external contaminants and from the determination of the detection efficiency.
This result is not competitive neither in terms of statistics nor in terms of energy threshold with other existing measurements of the $^8$B flux, but it represents an independent consistency check as it can be seen in Figure~\ref{fig:kl_8b_spectrum_cmp} (right) where all elastic scattering $^8$B flux measurements are compared with total errors. 
The weighted average is $\Phi_{\textrm{ES}}=2.33\pm0.05\times10^6\textrm{cm}^{-2}\textrm{s}^{-1}$ and is dominated by SK-I and SK-III results (section~\ref{sec:sk_III}).

\section{Future projects in Low Energy neutrino physics}
\label{sec:future}

The near future in solar neutrinos will see the start of the SNO+ detector, the conversion of the SNO detector into a liquid scintillator detector with a physics program that goes beyond low energy solar neutrinos including the search for 0-$\nu$ 2$\beta$-decay by loading of the scintillator with Nd or other candidate 2$\beta$-decay isotopes  \cite{zu11sno+}, detection of antineutrinos from the Earth (section~\ref{sec:geo}) and from reactors, neutrinos from the possible explosion of a galactic Supernova and searches for nucleon decay in "invisible" channels.
Many other projects are in different development stages like XMASS, MOON, CLEAN and LENS.

Physicists all over the world are already working since many years on next generation multipurpose detectors which are also very capable low energy neutrino detectors. 
Such detectors could be Megaton water Cherenkov detectors (Hyper-Kamiokande, Memphys, UNO), $\sim$ 50kton liquid scintillator detectors (LENA) of $\sim$ 5kton liquid Argon detectors (GLACIER).

For space reason we choose to briefly review here only SNO+ and LENA.

\begin{figure}[th]
\begin{center}
\includegraphics[width=.6\textwidth]{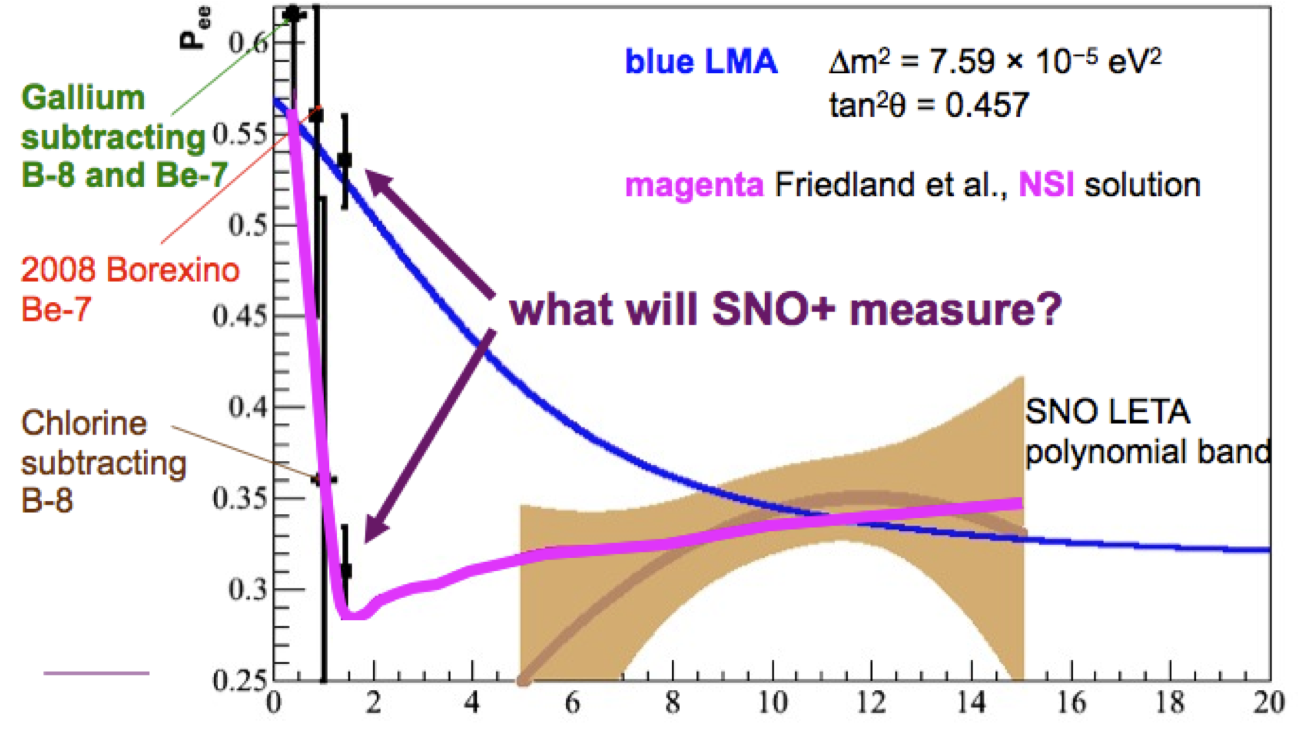}
\caption{Projection of SNO+ sensitivity for the $^8$B upturn. \cite{bil11sno+}.}
\label{fig:sno+}
\end{center}
\end{figure} 

\subsection{SNO+}
\label{sec:future_sno+}

SNO+'s idea is the conversion of the SNO detector (section~\ref{sec:sno_detector}) into an organic liquid scintillator detector \cite{bil11sno+}. 
The density of the scintillator in the central acrylic vessel is lower then the surrounding water, which reverts the situation of the heavy water target of SNO.
This posed the main construction challenge: the vessel support structure had to be replaced by a hold-down system, however a first pure scintillator phase is foreseen for early 2013 after which a 2$\beta$-decay phase is scheduled for a few years before pure scintillator is returned for the detailed solar study. 

In Figure~\ref{fig:sno+} is shown the simulated $^8$B spectrum compared to expectations from LMA and NSI models. 
As it can be seen by going below the energy range of the LETA analysis of SNO (section~\ref{sec:sno_leta}) it will be possible to distinguish the two models.
One of the dramatic advantages of this project is the very deep location, $\sim$6010m w.e. which virtually eliminates the problem of cosmogenic $^{11}$C background in the [1,2]MeV energy range. 
In SNO+ about 4000 events/yr are expected from pep neutrinos and a similar signal from CNO. 
The estimated error after three years of data taking is 5\% (pep) and 8\% (CNO) assuming the radio-purity levels obtained in Borexino (section~\ref{sec:bx}).

\subsection{LENA}
\label{sec:future_lena}

LENA is the proposed european next generation liquid scintillator experiment \cite{wu11lena}. 
Its physics program is broad, encompassing Low Energy physics (Supernova neutrinos from a galactic explosion but also the diffuse background; antineutrinos from reactors and from the Earth; solar neutrinos; indirect dark matter searches) and GeV Physics (proton decay, atmospheric neutrinos and long-baseline neutrino oscillations if targeted by a beam). 
Concerning solar neutrinos LENA could easily cover pep, CNO and $^7$Be neutrinos to unprecedented accuracy, but also go for the fundamental pp neutrinos in the lowest energy region.
The detector is the scale-up of present day experiments: 50kt liquid scintillator target in a cylindrical nylon vessel (13m radius, 100m height) surrounded by a buffer of 20kt inactive liquid and watched by 50000 8" PMTs for a 30\% optical coverage. 
A combination of Water Chrenkov and streamer tubes detectors would allow to track cosmic muons. 
The proposed location is a 4000m w.e. deep cavern in the Phyh\"asalmi mine (Finland). 
Research on this project started in 2005 (site studies, scintillator mixture development, PMT sealing, tank design, MonteCarlo, \ldots) and the collaboration based on 35 institutions from 12 countries has published the white paper in 2011 \cite{wu11lena}. 
Construction time is evaluated in eight years from the moment of approval.

\section{Geo-neutrinos}
\label{sec:geo}

Very little is known about the Earth interior. 
Neutrinos emitted by the Earth can escape freely and instantaneously and are therefore a powerful probe for the composition of the planet \cite{fio07geo}. 
In particular they can help to answer questions about the origin of the terrestrial heat (45TW), about the abundance of long lived radio-active U and Th in the crust and in the mantle and they can be used to test the hypothesis of a natural nuclear breeder reactor in the core of the Earth. 
The standard geophysical model for the description of the Earth is called Bulk Silicate Earth (BSE) and awaits to be tested by geo-neutrinos 

\begin{figure}[th]
\includegraphics[width=.45\textwidth]{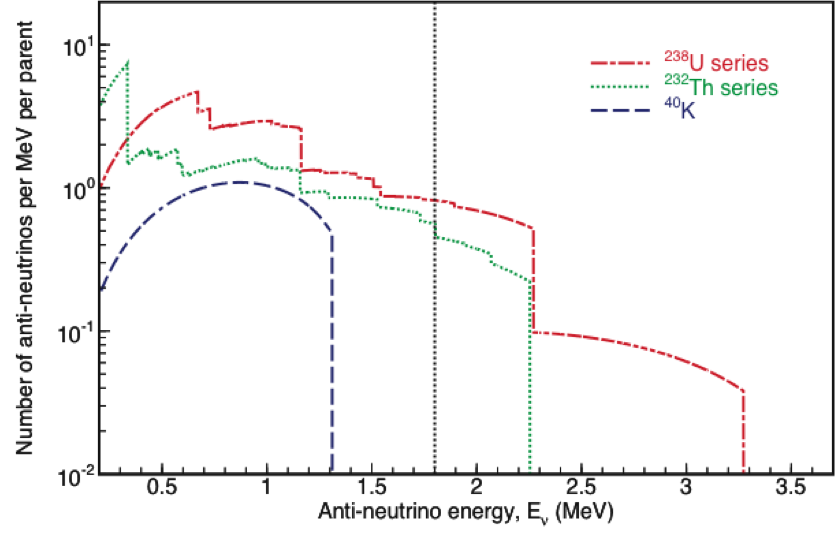}
\includegraphics[width=.53\textwidth]{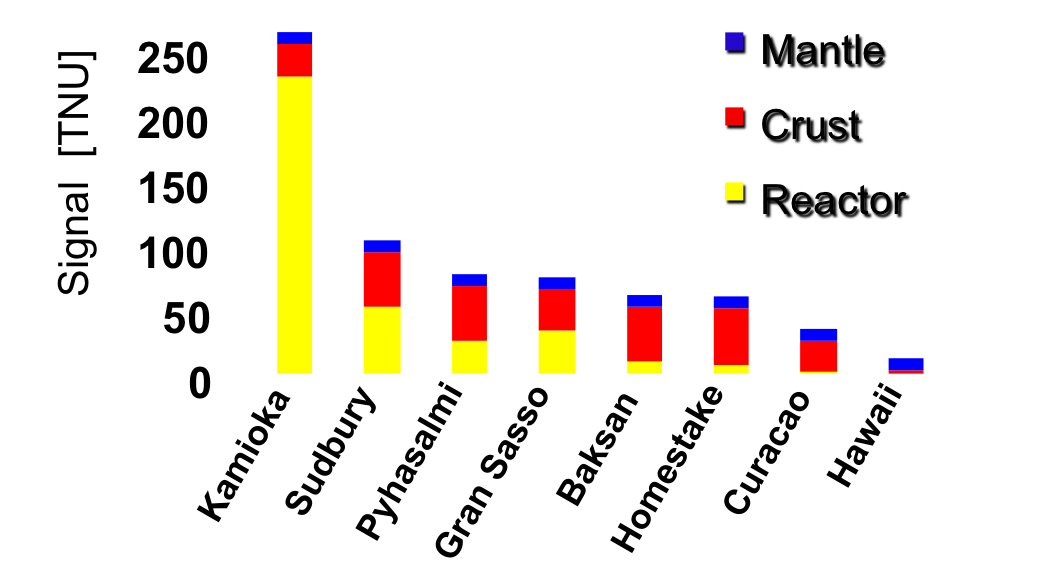}
\caption{Left: Energy spectrum of geo-neutrinos; the dashed vertical line shows the threshold for inverse $\beta$ decay. 
Right: Expected signal and background at current or planned sites for geo-neutrino detectors. \cite{fio07geo}}
\label{fig:geo_spectrum_sites}
\end{figure}

Geo-neutrinos are actually antineutrinos emitted within the decay chains of long lived radio-isotopes, namely $^{238}$U, $^{232}$Th 
and in the decay of $^{40}$K. 
The $^{238}$U chain emits neutrinos up to $\sim$3.3MeV maximum energy, while $^{232}$Th only reaches 2.2MeV and $^{40}$K  emits only $\overline{\nu}$ below the kinematic threshold for inverse beta detection (see section~\ref{sec:kl_reactors} for antineutrino detection in liquid scintillator detectors). 
The spectrum can be seen in Figure~\ref{fig:geo_spectrum_sites} (left).
The Th/U$\simeq$3.9 ratio is supposed to be fixed by the analysis of chondrites, meteorites that share origin with the solar system.
These elements are present in the crust although with unknown abundance, while it's not at all clear whether they are to be found in the mantle as well or not.

The oceanic crust is thinner ($<$10km) than the continental crust ($>$30km) consequently a detector located in the middle of a continent would see a higher signal then one located in a costal area and definitely higher then one in the middle of the ocean. 
At the same time a contribution from the mantle would be uniform. 
Consequently measurements performed by detectors "seeing" different crustal contributions could be combined to disentangle mantel and crust roles. 
On the other end the main background source for detection of geo-neutrinos are antineutrinos from nuclear power production (section~\ref{sec:kl_reactors}), so areas with a high density of plants are not good candidate locations.
In Figure~\ref{fig:geo_spectrum_sites} (right) the $S/B$ ratio of existing or planned experimental sites can be seen.

\begin{figure}[th]
\includegraphics[width=.49\textwidth]{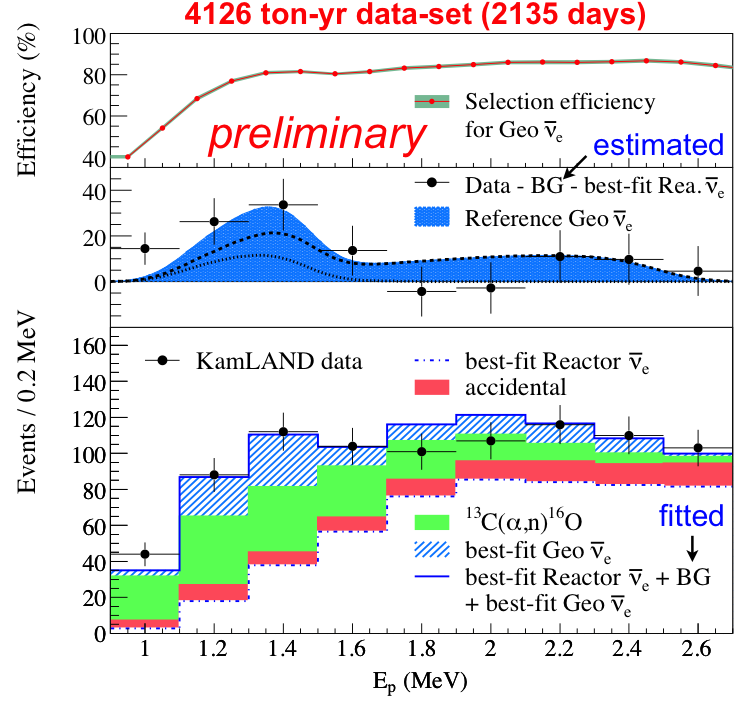}
\includegraphics[width=.49\textwidth]{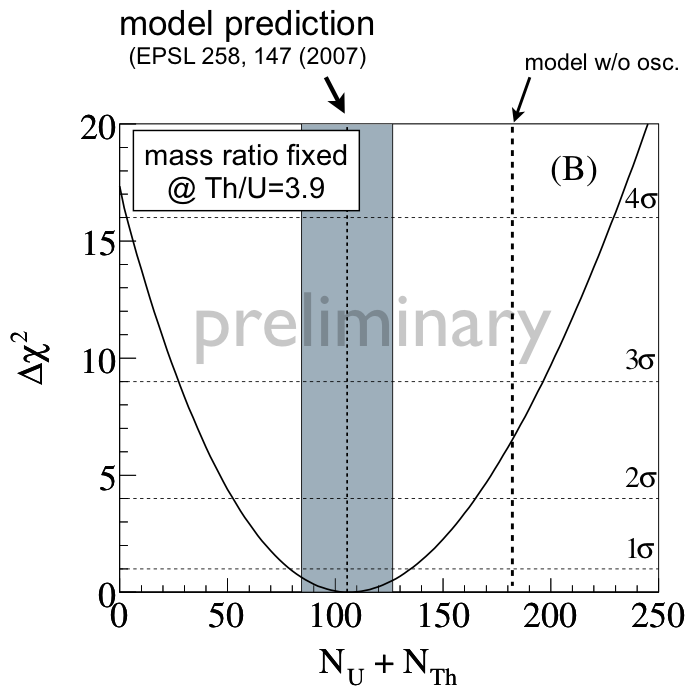}
\caption{KamLAND prompt energy spectrum for geo-neutrinos and the $\Delta \chi^2$ profile of the fit result \cite{kl10geo}.}
\label{fig:geo_kl}
\end{figure}

\subsection{KamLAND}
\label{sec:geo_kl}

The KamLAND experiment (section~\ref{sec:kl}) lies at the edge between continental and oceanic crust and in a region highly populated with nuclear power plants. 
The $S/B$ ratio is $\sim$0.2 without considering additional sources of background. 
Nevertheless they have analyzed a 4126 ton-yr data set for geo-neutrinos \cite{kl10geo} and found 841 candidate events in the [0.9,2.6]MeV energy range. 
Backgrounds have been evaluated to the following amounts: reactor $\overline{\nu_e}$ (484.7$\pm$26.5), $^{13}$C($\alpha$, n)$^{16}$O reactions (165.3$\pm$18.2), accidental coincidences (77.4$\pm$0.1), cosmogenic $^9$Li (2.0$\pm$0.1) and atmospheric $\nu$ plus fast neutrons (<2.8) for a total of 729.4$\pm$32.3. 
The hypothesis of no geo-neutrinos is therefore excluded from a rate-only test at 99.55\% C.L.
The energy spectrum is shown in Figure~\ref{fig:geo_kl} along with backgrounds and after their subtraction it is compared with expected signal from BSE.
The number of geo-neutrinos from a fit to the spectrum is $106^{+29}_{-28}$ assuming the fixed ratio U/Th=3.9.
The $\Delta \chi^2$ profile for the geo-neutrino component is also shown in Figure~\ref{fig:geo_kl}. 
The resulting flux is $\Phi_{\textrm{geo}}=4.3^{+1.2}_{-1.1}\cdot 10^6 \textrm{cm}^{-2}\textrm{s}^{-1}$.

\begin{figure}[th]
\includegraphics[width=.59\textwidth]{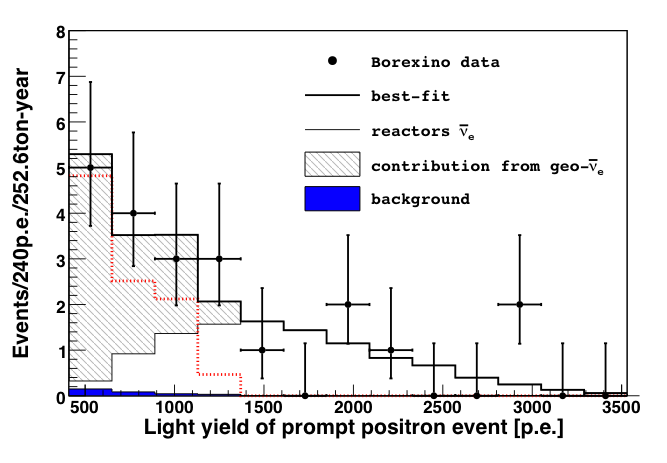}
\includegraphics[width=.39\textwidth]{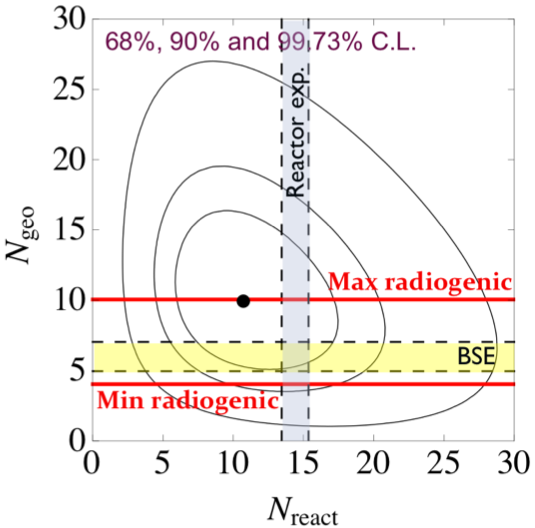}
\caption{Borexino prompt $\overline{\nu}$ energy spectrum and how the fit assigns the candidates to reactors and geo-neutrinos \cite{bx10geo}.}
\label{fig:geo_bx}
\end{figure}

\subsection{Borexino}
\label{sec:geo_bx}

Although smaller in size, Borexino is located on continental crust and about 800km from the nearest nuclear power plants. 
Its $S/B\sim0.4$ ratio is therefore more favorable then KamLAND's. 
In addition the much lower contamination of $^{210}$Po in  the scintillator (the $\alpha$ emitter responsible of the $^{13}$C($\alpha$, n)$^{16}$O reactions) and the thicker rock coverage make Borexino almost free from non reactor's backgrounds. 
The 252.6 ton-yr data set has been analyzed \cite{bx10geo} finding 21 anti-neutrino candidates. 
A fit to the energy spectrum (Figure~\ref{fig:geo_bx}) assigns $\sim$ 10 events to geo-neutrinos and $\sim$11 to reactor $\overline{\nu}$ for a resulting flux of $\Phi_{\textrm{geo}}=7.3^{+3.0}_{-2.4}\cdot 10^6 \textrm{cm}^{-2}\textrm{s}^{-1}$. 
The hypothesis of no geo-neutrinos is rejected at 99.997\% C.L. (4.2$\sigma$). 
Also the hypothesis of a geo-reactor at the Earth's center with more then 3TW power is disfavored at 95\% C.L.

\begin{figure}[th]
\begin{centering}
\includegraphics[width=.7\textwidth]{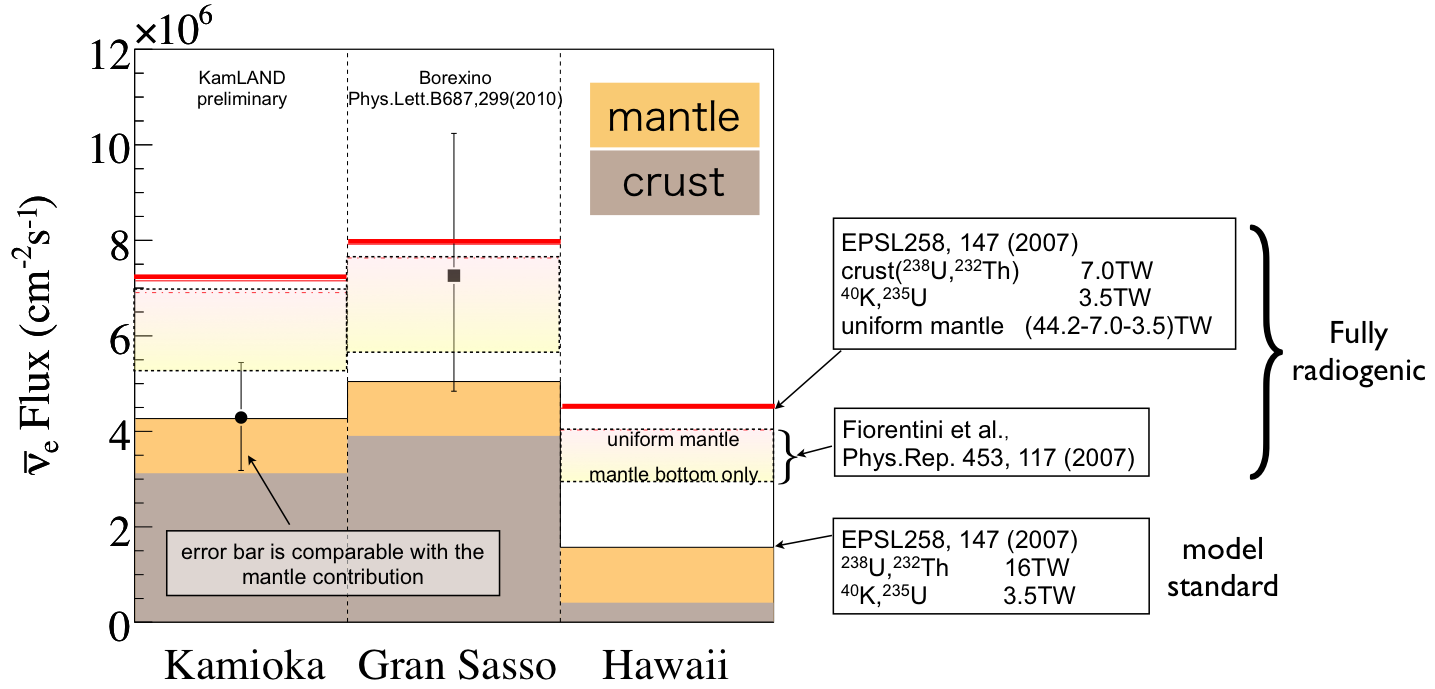}
\caption{Comparison between existing measurements of geo-neutrinos and predictions from BSE and fully-radiogenic models \cite{kl10geo}.}
\label{fig:geo_cmp}
\end{centering}
\end{figure} 

\subsection{Comparing data and model}
\label{sec:geo_cmp}

Figure~\ref{fig:geo_cmp} shows the fluxes measured by KamLAND and Borexino compared to geo-neutrino expectations from BSE at Kamioka mine and Gran Sasso, respectively. 
Mantle and crust contributions are evidenced. 
The BSE model assigns about 20TW of the terrestrial heat to radioactive decays, however alternative "fully radiogenic" models can ascribe all 45TW to this source. 
KamLAND result seems to favor the BSE model, while Borexino result is in a better agreement with fully radiogenic models, although the present day errors do not allow to draw conclusions yet. 
The third column in Figure~\ref{fig:geo_cmp} shows predictions for a proposed experiment (Hano-hano, \cite{dye06}) to be performed by sinking a liquid scintillator detector deep in the ocean. 
The experiment is proposed to take place on a custom ship located at the Hawaii islands, 
where the small crustal contribution would allow to measure geo-neutrinos from the mantle. 
The upcoming SNO+ experiment (section~\ref{sec:future_sno+}) thanks to its favorable location can count on a $S/B\sim1.2$ and is therefore expected to play a major role in the next years also in geo-neutrino physics.

\section{Conclusions}
\label{sec:conclusions}

Low energy solar neutrino experiments are giving fundamental contributions in answering the open questions in solar astrophysics and in understanding the mechanism of neutrino oscillations.
We have reviewed the recent progresses from Borexino: the experimental error on $^7$Be neutrinos has been reduced to to 5\% and no Day/Night asymmetry has been observed; $^8$B flux has been measured with the lowest threshold of 3MeV. 
The neutrino oscillation mechanism with MSW effect and the LMA solution has been therefore validated both in the low energy vacuum-dominated regime and in the high energy matter-enhanced regime, but a consistent effort is now put into the transition region ([1,3]MeV) where pep neutrinos are to be found and where the NSI model would show a different survival probability shape. 
To probe the transition region all experiments are trying to lower the detection threshold  of $^8$B neutrino. 
We have reviewed the evolution of Super-Kamiokande leading measurement through SK-III and SK-IV phases, we have reported about the powerful Low Energy Threshold Analysis from SNO that goes down to 3.5MeV with large statistics and about the consistency checks performed by KamLAND.
We have reviewed the latest reactor antineutrino data release from KamLAND and the global analysis combining them to solar data. 
Among upcoming projects we have selected SNO+ with great potentiality in lowering the $^8$B threshold to 1MeV and to detect pep and the so far undetected CNO neutrinos. 
Among next generation proposed experiments we have indicated LENA as very rich of potentiality in solar neutrino physics.

At the same time Borexino and KamLAND have measured the flux of antineutrinos produced by radioactive decays in the Earth's crust and possibly in the mantle. Such measurements are ideal probes of the geo-physical models such as BSE and more accurate measurements 
are awaited to understand how the terrestrial heat is generated and to shed light into the mystery of the Earth's composition.

\bibliographystyle{pramana}

\begin{thebibliography}{99}

\bibitem{msw}
L. Wolfenstein, PRD 17, 2369 (1978).
S.P. Mikheev and A. Yu. Smirnov, Sov. J. Nucl. Phys. 42, 913 (1985). 

\bibitem{pdg10}
K. Nakamura et al. (Particle Data Group), J. Phys. G 37, 075021 (2010).

\bibitem{sch11}
T. Schwetz, these proceedings.

\bibitem{SHP11}
A. M. Serenelli, W. C. Haxton and C. Pena-Garay, Astrophys. J. 743, 24 (2011).

\bibitem{Ser11nt}
A. Serenelli, Proc. of XIV "Neutrino Telescopes", Venice, Italy (2011). arXiv:1109.2602.

\bibitem{gs98}
N. Grevesse and J. Sauval, Sp. Sc. Rev. 85, 161 (1998).

\bibitem{ags09}
M. Asplund, N. Grevesse, J. Sauval and P. Scott, Ann. Rev. Astr. Astrophys. 47, 481 (2009).

\bibitem{ser09lowz}
A. Serenelli, S. Basu, J. W. Ferguson and M. Asplund, Astrophys. J. 705, L123-L127 (2009).

\bibitem{nsi04}
Friedland et al., PLB 597, 347 (2004).

\bibitem{bx08det}
G. Alimonti et al. (Borexino coll.), NIM A600, 568-593 (2009).

\bibitem{bx08be}
C. Arpesella et al. (Borexino coll.), PRL 101, 091302 (2008).

\bibitem{bx10B}
G. Bellini et al. (Borexino coll.), PRD 82, 033006 (2010).

\bibitem{bx11mv}
G. Bellini et al. (Borexino coll.) JINST 6, P05005 (2011).

\bibitem{bx11dn}
G. Bellini et al. (Borexino coll.), accepted for publ. on PLB (2011). arXiv:1104.2150.

\bibitem{bx11be}
G. Bellini et al. (Borexino coll.), PRL 107, 141302 (2011).

\bibitem{ca11met}
B. Caccianiga for the Borexino coll., proc. of 12$^{th}$ TAUP conference, Munich (2011). 

\bibitem{hol09}
P. C. de Holanda, JCAP 0907, 024 (2009).

\bibitem{sk03det}
Y. Fukuda et al. (Super-Kamiokande coll.), NIM A501, 418-462 (2003).

\bibitem{sk06I}
J. Hosaka et al. (Super-Kamiokande coll.), PRD 73, 112001 (2006).

\bibitem{sk10III}
K. Abe et al. (Super-Kamiokande coll.), PRD 83, 052010 (2011).

\bibitem{sk10IV}
S. Yamada for the Super-Kamiokande coll., talk at Physun (2010).
http://physun2010.mi.infn.it/PHYSUN\_yamada.pdf

\bibitem{sno99det}
J. Boger et al. (SNO coll.), NIM A449, 172-207 (2000).

\bibitem{sno07ph1}
B. Aharmin et al. (SNO coll.), PRC 75, 045502 (2007).

\bibitem{sno05ph2}
B. Aharmin et al. (SNO coll.), PRC 72, 055502 (2005).

\bibitem{sno11ph3}
B. Aharmin et al. (SNO coll.), subm. to PRC (2011). arXiv:1107.2901.

\bibitem{sno10leta}
B. Aharmin et al. (SNO coll.), PRC 81, 055504 (2010).

\bibitem{kl05rea}
T. Araki et al. (KamLAND coll.) PRL 94, 081801 (2005).

\bibitem{kl10cosm}
S. Abe et al. (KamLAND coll.), PRC 81, 025807 (2010).

\bibitem{kl11B}
S. Abe et al. (KamLAND coll.), PRC 84, 035804 (2011).

\bibitem{kl11rea}
S. Abe et al. (KamLAND coll.), PRD 83, 052002 (2011).

\bibitem{zu11sno+}
K. Zuber, these proceedings.

\bibitem{bil11sno+}
S. Biller, talk at XIV "Neutrino Telescopes", Venice, Italy (2011).

\bibitem{wu11lena}
M. Wurm et al., sub. to Astropart. Phys. (2011). arXiv:1104.5620.

\bibitem{fio07geo}
G. Fiorentini, M. Lissia and F. Mantovani, Phys. Rept. 453, 117-172 (2007).

\bibitem{kl10geo}
I. Shimizu for the KamLAND coll., talk at "Neutrino Geosience" (2010).
http://geoscience.lngs.infn.it/Program/Pdf\_presentations/Shimizu.pdf

\bibitem{bx10geo}
G. Bellini et al. (Borexino coll.), PLB 687, 299-304 (2010).

\bibitem{dye06}
S. T. Dye et al., Earth Moon Planets 99, 241-252 (2006).

%
%
\end{thebibliography}

\end{document}